\documentclass{sigcomm-alternate}
\usepackage[hang,flushmargin,stable]{footmisc}
\usepackage{amssymb}
\usepackage{amsfonts}
\usepackage{colortbl}
\usepackage{hhline}
\usepackage{color,verbatim}
\usepackage[bf]{caption}
\usepackage{indentfirst}
\usepackage{subfigure}
\usepackage{multirow}
\usepackage{tikz}
\usepackage{bbding}
\usepackage[all]{xy}
\usepackage[hyphens]{url}
\usepackage{times}
\usepackage{algorithmic}
\usepackage{array}
\usepackage{multirow}
\usepackage{balance}
\usepackage{paralist}
\usepackage{graphicx}
\usepackage{epstopdf}
\usepackage{amsfonts}
\usepackage{multirow}
\usepackage{subfigure}
\usepackage[bookmarks=false, colorlinks=true, plainpages = false,  linkcolor = darkblue,   citecolor = darkblue, urlcolor = darkblue, filecolor = darkblue]{hyperref}
\usepackage{verbatim}
\usepackage{ifthen}
\usepackage{xcolor}
\definecolor{darkblue}{RGB}{00,80,0}

\makeatletter
\def\url@leostyle{%
  \@ifundefined{selectfont}{\def\UrlFont{\fontsize{9}{10}\selectfont}}%
  {\def\UrlFont{\fontsize{9}{10}\selectfont}}%
}
\makeatother	
\newcommand{\descr}[1]{\vspace{0.2cm} \noindent \textbf{#1}}

\newcommand{\changed}[1]{\textcolor{black}{#1}}

\newcommand{\shortversion}{false} %
\newcommand{\shortver}[1]  {\ifthenelse{\equal{\shortversion}{true}}{{#1}}{}}
\newcommand{\longver}[1]  {\ifthenelse{\equal{\shortversion}{false}}{{#1}}{}}

\title{Privacy in Content-Oriented Networking:\\ Threats and Countermeasures{\LARGE\thanks{~~{\em A preliminary version of  this paper appears in the ACM SIGCOMM Computer Communication Review (CCR), Vol. 43, Issue 3, July 2013. This is the full version.}}}}

\author{Abdelberi Chaabane{\large$^1$}, Emiliano {De Cristofaro}{\large$^2$}, Mohamed Ali Kaafar{\large$^{1,3}$}, and Ersin Uzun{\large$^2$}
\vspace{0.35cm}\\
\begin{tabular}{ccc}
\fontsize{10}{10}\selectfont \sf $^1$ INRIA Rhone-Alpes & 
\fontsize{10}{10}\selectfont \sf $^2$ PARC & 
\fontsize{10}{10}\selectfont \sf $^3$ NICTA\\
\fontsize{10}{10}\selectfont \sf Montbonnot, France & 
\fontsize{10}{10}\selectfont \sf Palo Alto, CA, U.S.A. & 
\fontsize{10}{10}\selectfont \sf Sydney, Australia
\end{tabular}
}

\begin{document}
\maketitle
\thispagestyle{plain}
\pagestyle{plain}

\begin{abstract}
As the Internet struggles to cope with scalability, mobility, and security issues, new network architectures are being proposed to better accommodate the needs of modern systems and applications.  In particular, Content-Oriented Networking (CON) has emerged as a promising next-generation Internet architecture: it sets to decouple content from hosts, at the network layer, by naming data rather than hosts. CON comes with a potential for a wide range of benefits, including reduced congestion and improved delivery speed by means of content caching, simpler configuration of network devices, and security at the data level. However, it remains an interesting open question whether or not, and to what extent, this emerging networking paradigm bears new privacy challenges. In this paper, we provide a systematic privacy analysis of CON and the common building blocks among its various architectural instances in order to highlight emerging privacy threats, and analyze a few potential countermeasures. Finally, we present a comparison between CON and today's Internet in the context of a few privacy concepts, such as, anonymity, censoring, traceability, and confidentiality. 
\end{abstract}

\section{Introduction}
\label{sec_intro}

In the last few years, the increasing penetration of the Web in our society 
has prompted a tremendous growth of data routed in the Internet, to such an extent that
global IP traffic\longver{-- thrusted by the soaring proliferation of mobile 
and streaming data --} is expected to increase 3-fold over the next 5 years~\cite{cisco}.
Besides exceeding its expectations, the Internet has also stretched many of the initial assumptions,
creating issues that challenge its underlying communication model. 
Applications increasingly operate in terms of content, making it difficult to conform to
IP's requirement to communicate by discovering and specifying hosts and locations~\cite{ndn}.

Coping with massive amounts of traffic becomes arduous due to a number of
issues deep-rooted in the network design.
As a result, the quest for improving scalability and (cost) efficiency of content delivery 
has led to the design of overlay networks, such as, Peer-To-Peer (P2P) and Content Distribution Networks (CDNs). 
However, overlay networks often complicate network management and application development.
\longver{P2P reduces servers' load by distributing content among peers,
using dynamic and fault-tolerant networks, but it often results in an increased inter-provider traffic~\cite{bt01,bt02}. 
Moreover, P2P is application-dependent, thus, confined to a specific usage. 
CDNs are used to convoy user's requests
to geographically closest caches in order to reduce traffic, however, 
they require ad-hoc infrastructures and only some providers can afford the related deployment costs. 

} %
Further, endpoint authentication mechanisms (whereby an endpoint can only authenticate the counterpart, 
but not the message) have been challenged by 
frequent attacks against SSL~\cite{ssl_flow,georgiev2012most} and the hacking of certification authorities~\cite{ssl_hack}.
\longver{Also, the Internet today often struggles with mobility and resilience to disruption. Transport layer is, by design, unable to manage mobile parties and add-on features -- e.g., Mobile IPv6 (MIPv6) and Hierarchical MIPv6 \cite{hmip} -- have been suggested, albeit suffering from handoff latency and packet losses \cite{mobile_ip}.
} %
   
Motivated by these issues, new architectures have been proposed, in the last few years, aiming to redesign the Internet 
\longver{(see, e.g., NSF's Future Internet Architecture multi-million program~\cite{NSF}),}
\shortver{\cite{NSF}}
and accommodate content-oriented applications. In particular, \textbf{Content-Oriented Networking (CON)}~\cite{cho2008content} has set
to decouple contents from hosts, at the network layer, by relying on the publish/subscribe paradigm. %
CON shifts identification from host to content, so that this can be located anywhere in the network.
The content-centric communication paradigm introduced by CON relies on naming the content itself, rather than its location\shortver{.}\longver{, and thus radically changes the way data is handled.} Content is self-contained, has a unique name, can be retrieved by means of an \textit{interest} for that name, cached in any arbitrary location,
and digitally signed to ensure its integrity and authenticity.

\longver{\subsection{Roadmap \& Contributions}}
CON comes with a potential for several advantages, including reduced congestion and improved
delivery speed through content caching, simpler configuration of network devices, and 
security at the data level.  However, it remains an interesting open question whether or not, and to what 
extent, this emerging networking  paradigm bears new \textbf{privacy challenges}. 
While some features, such as, lack of source/destination addresses, might help privacy in CON, 
a closer look to some of the design choices unveils a number of open questions. This paper \longver{systematically}%
studies privacy in CON as a generic paradigm and shows that it introduces several worrisome issues.

First, we analyze the implications of {\em caching} 
-- one of the crucial features in CON used to reduce traffic and improve delivery speed -- 
on user privacy and show that, as nodes cache frequently requested content, they can infer content consumed by others using timing information. 
Next, we focus on {\em content privacy}: since publicly available content is not encrypted in CON 
and, as routers now handle content, any router can easily inspect content.
\longver{Whereas, packet inspection in today's Internet requires routers to reassemble packets,
which is very inefficient as router primary job is to forward packets, and often requires dedicated hardware~\cite{dpi}.

}%
Furthermore, as content in CON is retrieved using names that most likely are semantically related
to the content itself, an attacker could infer sensitive information about a user, by monitoring her requests: we refer to
this issue as {\em name privacy}.
Finally, we look at {\em signature privacy}: since digital signatures of CON packets need to be publicly verifiable, the identity of a content signer may be easily inferred by looking at the signature.

In this paper, we discuss different attack scenarios that threaten privacy in CON. 
For each setting, we describe the attacker capabilities and related impact on user privacy. We suggest several \changed{ideas that could serve as countermeasures} and detail  their strengths and weaknesses, and make sure that they would bear minimal changes to the CON architecture. To the best of our knowledge, this represents the first step toward a thorough analysis of CON privacy issues.
In the process, we also highlight a number of challenging open problems that call for further research.

\longver{
\descr{Paper Organization.}
Next section presents an overview of Content-Oriented Networking (CON). Then, we provide a thorough analysis of a few privacy challenges in CON and  detail possible countermeasures in Section~\ref{sec:PrivacyChallenges}.
In Section~\ref{sec:Discussions}, we compare the feasibility and effectiveness of privacy-enhancing technologies in CON as opposed to today's Internet. Finally after summarizing related work in Section~\ref{sec:related}, the paper concludes in Section~\ref{sec:conclusion}.
}

\section{Content-Oriented Networking}
\label{overview}

This section provides a high-level description of Content-Oriented Networking (CON).
\longver{We present the main components common to most of available CON architectures, along with
their respective design choices.

}%
A few future Internet architectures have been proposed so far that realize CON. The most prominent ones 
include DONA~\cite{dona}, NetInf~\cite{netinf}, CCN~\cite{ccn}, LANES~\cite{psirp}, TRIAD~\cite{triad}, CBCB~\cite{cbcb}.
We now review their \longver{macro}%
building blocks:\vspace{0.1cm}

\begin{compactenum}
\item {\bf Named content:} In CON, objects are always named to facilitate data dissemination and search. Consequently, the security model is also shifted from host to content authentication. 
\vspace{0.1cm}
\item {\bf Content-based routing:} Content routing in CON relies on content rather than hosts, aiming to handle increased amounts of network traffic and be more resilient to network bursts and users' mobility.
\vspace{0.1cm}
\item {\bf Content Delivery:} Content is efficiently delivered using multi-path routing and leveraging in-network caching, in order to minimize network bandwidth and delivery delay, and transparently handle mobile users.
\vspace{0.1cm}
\item {\bf In-network storage:} All CON network components provide caching capability.
Note that this is different from packet buffers in today's routers, as cache size is expected to be several orders of magnitude bigger in CON.
\vspace{0.1cm}
\end{compactenum}

\begin{figure}[t!]
  \begin{center}
  \includegraphics[scale=0.43]{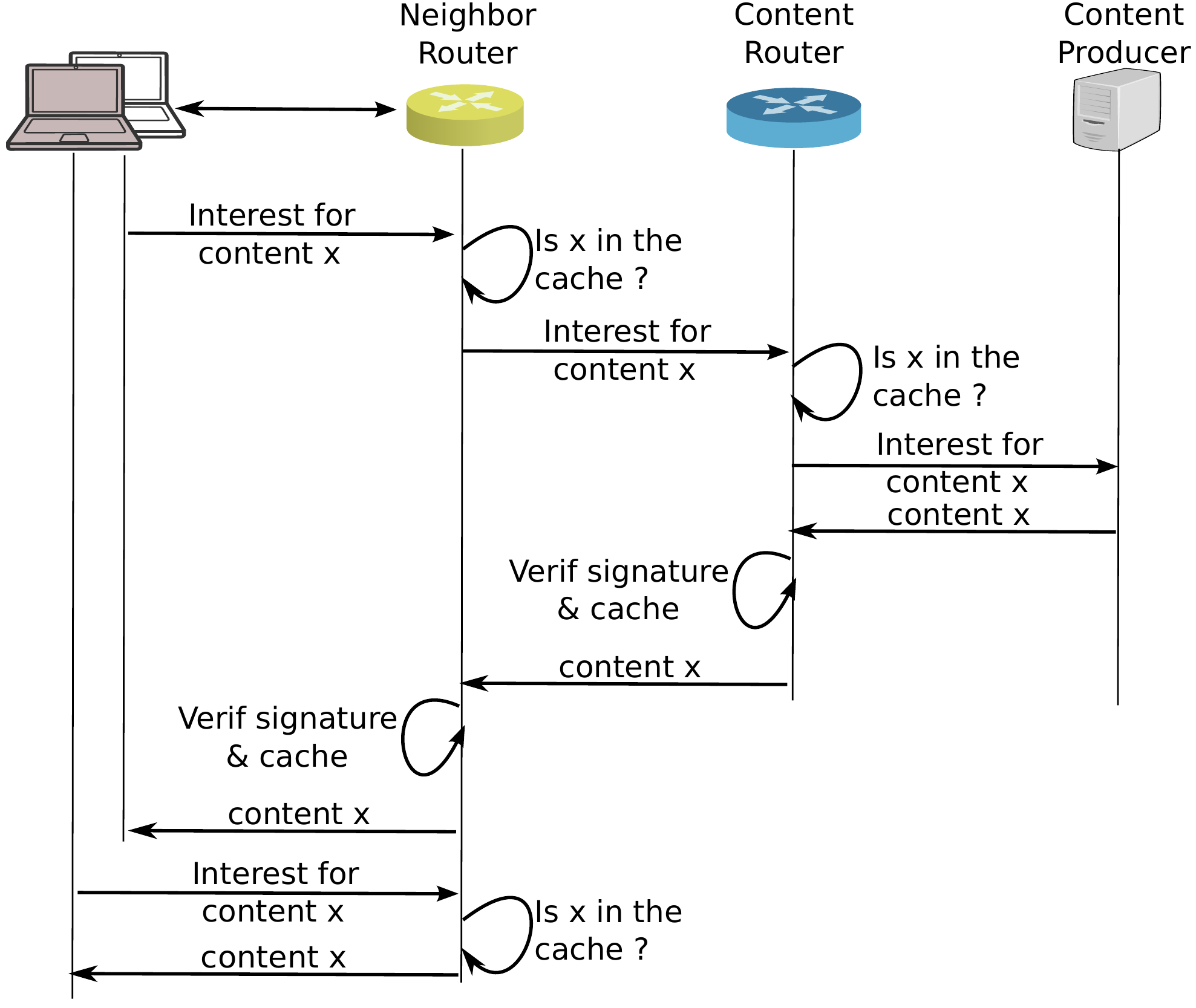}
  \caption{An overview of the main CON features: content routing, caching, and content signature. Content is address by name (x).}
  \label{fig:overview}
	\end{center}	
\vspace*{-0.4cm}
\end{figure}

\descr{Network Model.}
CON involves several entities (see Fig.~\ref{fig:overview}). {\em End users} express interests and fetch data using a wide range of devices. Interests represent the willingness of the user to retrieve certain data, independently of its location. {\em Content routers} are responsible of forwarding interests and forwarding back the associated data. Each router is assumed to  have a built-in cache. Cache size as well as caching algorithm may differ from one router to the other. Finally, {\em content producers} (or publisher) generate the content---either static (time-independent) or dynamic (generated upon request). Although data-centers or/and geographically distributed servers may be used to serve content, we simplify the model by considering a single source machine.

\smallskip
In the following, we overview CON's architecture design. For further details, we refer the reader to~\cite{survey_ccn_01,survey_ccn_02}\longver{.}%
\shortver{, and to the full version of this paper~\cite{full}.}

\descr{Caching.}
A key feature used to increase overall network efficiency in CON is caching. All nodes in the network are expected to participate in the caching effort, from core routers to mobile devices. Caches provide in-network storage and are assumed to be several orders of magnitude larger than today's buffers. This capacity allows to store content for longer periods and enhance network performance. 
There is a number of efforts aiming at optimizing caching strategies in CON -- see, e.g.,~\cite{aggregation_effect,probabilistic_caching,caching_perf02,caching_perf}. 
\longver{In particular, \cite{caching_perf} demonstrates that  network topology has a limited effect in caching efficiency, whereas, catalogue's size (i.e., size of all content)  and content popularity play a major role in caching efficiency.  \cite{probabilistic_caching} proposes a probabilistic caching algorithm, based on an approximation of the cache capacity, while~\cite{wave} introduces a caching approach that relies on the content popularity to decide whether to cache the content or not. }

\descr{Content Naming.}
\label{sec:naming}
The main abstraction in CON architecture is the Content Object (CO). All kinds 
of content, ranging from web pages to documents, and even interactive content, such as VoIP, are abstracted as a CO. An object is always identified by a name, which must be unique as it serves as identifier for searching and disseminating associated content. Moreover, since content can be fetched from anywhere, there should be a \emph{secure binding} between content name and content data (i.e., \textit{name-data integrity}) as well as \textit{object authenticity}. Finally, objects retrieved from cache should carry information about the object owner (publisher). Two naming approaches have been proposed so far -- flat and hierarchical.

\descr{Flat naming.}  Flat names are \textit{self-certifying}, i.e., CO's \textit{name-data integrity} is bounded to its name, thus, it can be verified without any PKI. In its simplest form, the name is expressed as the cryptographic hash function of the content. Although one can assess the validity of the content, it is impossible to verify its provenance and relevance. However, a few techniques have been proposed to enable provenance verification, by allowing the publisher to have more control on the naming by adding a label (e.g.~\cite{OceanStore, dona}). Unfortunately, these solutions do not guarantee binding between the name and the content. Thus, we will assume the use of secure binding, as proposed by~\cite{pracName}, whereby content is made available as a triple $M_{(N,P,C)} = (N,C, Sign_{P}(N,C))$, where $N$ is the content name, $C$ the content, and $Sign_{P}$ producer's signature.

\longver{
For instance, the naming mechanism can be instantiated as $Digest(PublicKey_{p} || Label_{c})$~\cite{OceanStore}, where $||$ denotes concatenation and $PublicKey_{p}$  is the content producer public key, or $Digest(PublicKey_{p}) || Label_{c}$, as it happens in DONA~\cite{dona}. The content is authenticated by checking whether or not it is verifiable using $PublicKey_{p}$. Unfortunately, both solutions do not guarantee binding between the name and the content.\footnote{\scriptsize We will assume the use of secure binding, as proposed by~\cite{pracName}, whereby content is made available as a triple $M_{(N,P,C)} = (N,C, Sign_{P}(N,C))$, where $N$ is the content name, $C$ the content, and $Sign_{P}$ producer's signature.} Moreover, as pointed out by~\cite{pracName}, flat naming suffers from several other disadvantages. First, opaque names are location-free, thus, making it difficult to build a routing mechanism to retrieve the nearest copy \cite{flatLabel,lisp-dht}. Therefore, location-dependent approaches are often used, such as Distributed Hash Table (DHT) \cite{OceanStore,sdsi,tapestry}. Second, and most importantly, it is well-known that users strive to remember simple strings like emails and hostnames~\cite{internationalNaming,User-Relative-Names,PersoNamespaces}, 
thus, indirection architectures~\cite{internationalNaming,layered-naming} become necessary. This is somewhat similar to the Domain Name System (DNS) service (used today to map user-friendly names to network names), and, unfortunately, shifts security/privacy problems, once again, from the content name to the mapping architecture. 
}

\descr{Hierarchical naming.} With hierarchical naming, name structure resembles that of today's URLs, where the '/' symbol delimits the name components. %
In some cases (e.g., CCN~\cite{ccn} and TRIAD~\cite{triad}), the name is human-readable,  which makes it easier for the user to access the content. The major benefit of adopting a hierarchical structure is to enable aggregation, thus, improving routing scalability. \changed{Note that, similar to flat naming, we assume the use of secure binding, as proposed by~\cite{pracName}.}

\longver{Nonetheless, hierarchical naming has at least three drawbacks. First, names are not persistent: any changes in the hierarchy alter the content name and hence make the content unreachable. Second, since names are semantically meaningful, it might leak sensitive information about the content and hence may be considered as a privacy threat (which we further explore in Section~\ref{sec:signature}).
Finally, a Public-Key Infrastructure (PKI) is needed to verify name-data integrity, potentially impacting the scalability and the security of the architecture.}

\descr{Content Routing and Forwarding.}
As illustrated in Fig.~\ref{fig:overview}, routing in CON is performed in two phases: (1) routing of CO requests (``interests''), and (2) routing the content back to the user.
Naturally, this depends on the naming schema and, in particular, on whether or not name aggregation is possible.
For flat-naming based CON, a Name Resolution Service (NRS) is used to retrieve topological information (such as, the data location) based on object name.
\emph{Structured routing} algorithms are often used to exploit structured network topologies, such as trees 
or DHT-s. For example, DONA~\cite{dona} maintains a tree topology and lets each router store routing information of all his descendants. Thus, any content (re)publication, deletion, or modification is propagated up to the root. 
With hierarchical naming, efficient routing and discovery is possible without any external service. Request/data aggregation may also facilitate network scalability. This approach may resemble unstructured routing in IP, where IP addresses are  replaced with content name and route advertising is achieved through flooding. 

\descr{CCN/CCNx overview.} 
To ease presentation, in the rest of the paper, we will refer to technical details of one specific CON instance, namely, CCNx~\cite{ccnx}, the open-source project that implements Content-Centric Networking (CCN)~\cite{ccn}.
\changed{CCNx is considered one of the most mature examples of CON in the research community, with multiple platform implementations, documentations, and implementations, supported by a relatively large community.}

In CCNx, whenever a router receives an interest for name $X$, it performs a longest prefix match lookup on its three main tables as follow: First, it looks whether the interest exits in the Content Store (i.e., the main cache): if so, a copy is forwarded back to the user and the routing process terminates. Otherwise, a lookup is launched on the Pending Interest Table (PIT). This table keeps track of all interests waiting to be resolved by upstream nodes. If there is a match,  the router collapses the present interest (and any subsequent ones for $X$) storing only the interface on which it was received. Finally, if no match is found, the router searches for the most suitable interface in his Forward Information Base (FIB) to forward the interest and then creates a PIT entry for that interest.   

\section{Privacy Challenges in CON}
\label{sec:PrivacyChallenges}
We now present a systematic analysis of privacy in Content-Oriented Networking (CON) by identifying threats and, when possible, discussing possible solutions. Multiple proposals~\cite{dona,netinf,ccn,psirp,triad,cbcb} 
have been presented in the last couple of years to instantiate CON, with relatively minor differences in their proposed design. 
\changed{As mentioned above, we will discuss technical details while considering the CCNx instance~\cite{ccnx},
however, threats and proposed countermeasures discussed in this paper apply to CON in general.
That is, issues are about fundamental features in CON (such as, caching, naming, data delivery and provenance assurance) and not about the specifics of one implementation. }

\subsection{Cache privacy}
\label{sec:CachePrivacy}

\longver{
\begin{figure}[t!]
  \begin{center}
  \includegraphics[scale=0.4]{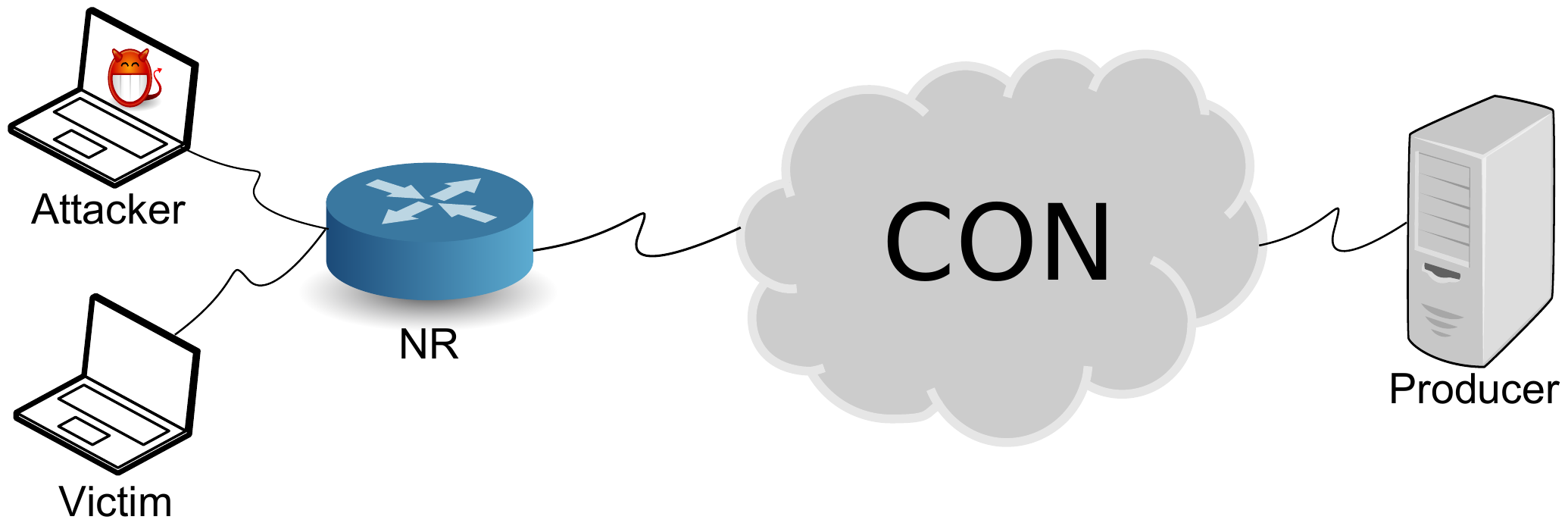}
  \caption{Topology of cache attacks.}
  \label{fig:cache_attack}
	\end{center}	
\end{figure}
}
We start our analysis with cache privacy. Recall that caching is a fundamental component of CON as 
it benefits both latency and bandwidth consumption. 
However, it also introduces a fundamental challenge to user privacy. An adversary may use a router's cache to infer content  exchanged (consumed) by users in the downstream and possibly link it to a specific user depending on its relative location to the user in the topology. 
\longver{Cache attacks in CON are exemplified in Fig.~\ref{fig:cache_attack}.}
There are different types of attacks that can occur on cache privacy -- we review them below.

\descr{Timing attacks.} By measuring time~\cite{timing_attack}, an adversary $Adv$ can determine if a content has been cached at a particular router by measuring the delay to retrieve it. To do so, $Adv$ measures the $RTT_{s}$ to retrieve any content from the source, the delay $RTT_{c}$ to get cached content from the closest router, and the delay $RTT_{t}$ to fetch targeted content. Then, $Adv$ compares the $RTT$ as follow:
\begin{compactitem}
\vspace{0.1cm}
\item If $|RTT_{t} - RTT_{c}| < \epsilon$ (for negligible $\epsilon$): $Adv$ concludes that target content has been cached at the closest router (i.e., has been fetched by a neighboring consumer connected to the same router).
\vspace{0.1cm}
\item If $RTT_{t} > RTT_{c}$ and $RTT_{t} < RTT_{s}$: $Adv$ knows that target content has been fetched from the source recently and cached in the network, but not by one of its immediate neighbors. Based on the difference between $RTT_{t}$ and $RTT_{c}$, $Adv$ can still predict how close the consumer of that content was to his location in the network topology.
\vspace{0.1cm}
\item Otherwise ($|RTT_{t} - RTT_{s}| < \epsilon$), $Adv$ concludes that target content has not been consumed recently.
\vspace{0.1cm}
\end{compactitem} 
Such an attack allows $Adv$ to check whether or not content has been recently fetched, but not when. \changed{Launiger et al.~\cite{tobi_2} describe the timing attacks in the context of CON, and show that the cache replacement policy has only a small impact on the attacks' success.} Finally, observe that, since in-network caches are shared by all CON designs, this attack is inherent to \textit{all} CON proposed architectures.

\descr{Protocol attacks.} Without a careful design, content retrieval protocols and their features in CON architectures can make access to cache content even easier. After investigating such issues in CCNx~\cite{ccnx}, we found out that a number of features and options in interest packets~\cite{ccn_interest}, and how they are matched with content packets, are particularly worrisome:
\begin{compactitem}
\vspace{0.1cm}
\item \emph{Prefix-based matching}: CCNx considers a content with name $X$ to satisfy an interest for name $Y$ if $Y$ is a proper prefix of $X$. This can facilitate easy extraction of cache content without knowing exact names. Due to multiple types of content potentially satisfying an interest, an exclusion option is also conveniently provided in CCNx interest packet format to allow exclusion of previously acquired content from subsequent queries.
\vspace{0.1cm}
\item \emph{Scoping}: In CCNx, scope for interest packets is used to determine the maximum number of hops it will travel. Such a feature makes it easy to query the caches of particular routers as it controls where (i.e., how many hops away) an interest packet can travel to. For instance, setting the scope to 2 would restrict an interest to propagate to only neighbouring router(s) and allow convenient querying of their caches without relying on any timing information.
\vspace{0.1cm}
\end{compactitem} 
As a result, $Adv$ can \emph{monitor} the access to sensitive content within a certain scope or easily \emph{dump} nearby caches' content. The former attack is achieved by periodically issuing an interest $I_{m}$ for target content $m$ and setting the scope accordingly (e.g., setting scope to 2 to monitor the caches of immediate neighbors). If $I_{m}$ times out, $Adv$ concludes that $m$ has not been fetched yet. Whenever a consumer within the scope accesses $m$, $I_{m}$ is satisfied, thus, allowing $Adv$ to be notified. Hence, $Adv$ is able to infer \textit{when} the file was fetched for the \textit{first} time. 
Dumping attacks can be achieved by sending an interest for the root prefix $/$ or short prefixes, repeatedly, and excluding what has been already received on successive interests. Combined with scoping, this method can easily be used to dump cache contents from nearby caches.

\smallskip
We believe that the above attacks are quite worrying, however, observe that the attack success depends on the relative location of the adversary to the victim in the network topology. Thus, we distinguish two classes of adversaries:
\begin{compactitem}
\vspace{0.1cm}
\item {\em Immediate Neighbor:} if an attacker is sharing the first hop CON router with his potential victim, the privacy risk is maximized as it would not only be easy to singularly monitor or dump a close-by router, but also victim's anonymity set would be very small, due to the limited number of users sharing that router. 
\vspace{0.1cm}
\item {\em Distant Neighbor:} Considering the tree-like topology in content distribution from its original source to its consumers (where the source is the root, consumers are the leaves, and the intermediate routers are nodes in between), the path from an adversary and a consumer to the root will intersect at least one node. Therefore, the privacy risk decreases as the number of leaves in the sub-tree rooted at that node increases (i.e., anonymity set gets larger).
\vspace{0.1cm}
\end{compactitem}

\subsubsection*{Potential Solutions}

Different algorithms have been proposed to enhance the hit ratio~\cite{aggregation_effect, probabilistic_caching,caching_perf02,caching_perf} on caches, however, none of them takes into consideration potential related privacy issues. In the following, we discuss some potential countermeasures to mitigate such privacy threats at a high level. Although the detailed design and security analysis for these methods are not within the scope of this paper, we expect the research community to further investigate them in future work.

\descr{Wait before reply.} A simple solution to the cache privacy problem is to delay \emph{all} requests: when the router fetches content $m$, it should store the corresponding RTT $t_{m}$. 
Then, whenever a user requests $m$, the router waits $t_{m}$ before sending the data back. 
\changed{Note that, independently from our work, Acs et al. have recently proposed a similar solution~\cite{acs}.}
Wait before reply has three main advantages: (1) \changed{it provably achieves perfect privacy since $Adv$ cannot distinguish between cached and not cached data~\cite{acs}}; (2) it does not make any assumptions about content correlation, network topology, or consumers, and (3) it achieves reduced bandwidth thanks to caching. 
Unfortunately, however, this approach has the main drawback of eliminating the positive effect of caching on content retrieval delay. 

\descr{Delay the first \textit{k}.} 
An alternative to the above solution is to delay the first $k$ requests for content $m$ to ensure that only popular content is cached on edge routers serving small number of customers,
similar to what Acs et al. have recently (and independently from our work) proposed in~\cite{acs}.
Note that $k$ should be chosen randomly by routers, otherwise an adversary could break this schema by issuing $k$ requests and timing responses. The main advantage of this approach is that consumers accessing popular content are unlikely to experience any delays introduced by routers. 
However, $k$ should be carefully and randomly chosen for each content. High values of $k$ result in delaying most of the requests, whereas, a small value will have a negative impact on user's privacy by reducing the anonymity set. Furthermore, the delay on the retrieval of not so popular content will still be high. \changed{~Finally, we note that Acs et al.~\cite{acs} provide a formal model that allows to quantify the tradeoff between privacy offered by various caching algorithms and the latency.}

\longver{
\begin{figure}[t!]
  \begin{center}
  \includegraphics[scale=0.5]{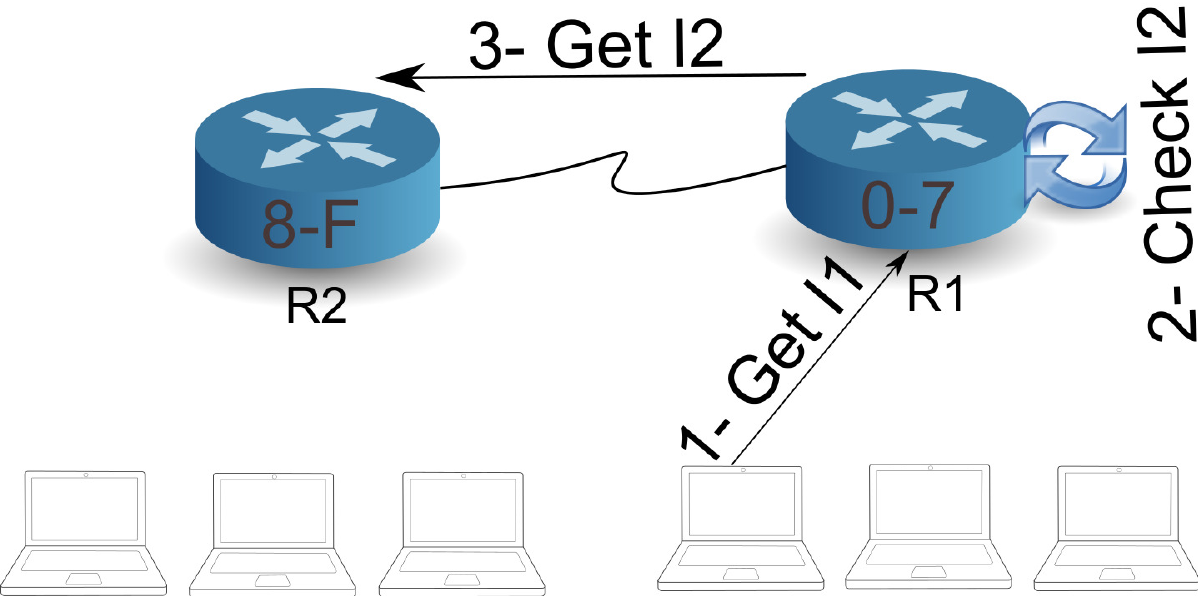}
  \caption{Collaborative caching}
  \label{fig:collCaching}
	\end{center}	
\end{figure}
}
\descr{Collaborative caching.} Multiple nearby caches could collaborate to create a distributed cache that is serving a bigger set of users. Such an integration would create an illusion of a single cache of bigger size and would also increase the anonymity set for customers. Several algorithms have been proposed in this context~\cite{h-cache,internet_cache} and can be categorized as hierarchical or mesh based. The latter refers to a flat structure while the former is used for caches that have a tree-like structure. In a simple mesh scenario, two routers $R_{1}$ and $R_{2}$ collaborate as follows: the hash universe (e.g., 256 bits) is divided in two subspaces $s_{1}$,$s_{2}$ where $R_{1}$ and $R_{2}$ stores the elements in $s_{1}$ and $s_{2}$ respectively. Based on computed hash (of the routable prefix of the content name), the router decides whether to cache that content or transmit it to his neighboring router. Similarly, interests are first forwarded to the router responsible for caching the corresponding subspace\longver{(see Fig.~\ref{fig:collCaching})}.
Collaborative caching has two main advantages: first, users are likely to fall into larger anonymity sets even if the requested content was found in cache. Second, hit rates for caches will increase as the collaboration would remove redundancy between nearby caches and effectively simulate a cache that is much bigger in capacity. Due to this second property, there may be some economic incentive to deploy this solution besides protecting user privacy as well \cite{squid_cache}.

\descr{Probabilistic caching.}  Introducing randomness in the caching procedure may impact the accuracy of attacks. One possible approach could be probabilistic caching~\cite{probabilistic_caching},
where a router decides to cache content based on his position on the forwarding path as well as the available space in the cache. Since this decision is based on internal states
of routers, it would not be known to an adversary. However, not caching a random subset of content can provide only a very limited privacy protection as the cached subset would still violate user privacy.

\subsection{Content privacy}
\label{sec:content}

Unencrypted communication over IP networks can be spied upon using Deep Packet Inspection (DPI)~\cite{dpi} by an adversary on the end-to-end communication path. However, in CON, content privacy becomes an even more serious threat due to the presence of persistent memory (caches) within the network.

\descr{Monitoring and Censorship:} DPI tools are already commonly used by certain governments or Internet Service Provider (ISP) for classifying and censoring content (e.g., based on keywords). However, DPI on IP networks requires powerful adversaries that are strategically located on the main communication path and with enough computation power to perform DPI on line speed.
As CON stores data packets for long time and makes it available to anyone that asks for it, neither of these assumptions about the adversary
holds. In fact, the adversary might retrieve content from caches for DPI based monitoring, classification and censorship. 

\subsubsection*{Potential Solutions}

As the problem is caused by the lack of data confidentiality, encryption would be the de-facto solution. Naturally, the encryption mechanism
should provide the best balance between security and efficiency. Also, it should be preserving the benefits of caching mechanism.
 
\descr{Symmetric/Asymmetric encryption.}  A trivial approach might be to use similar mechanisms to SSL/TLS, where a client %
generates a session key and encrypts it using the producer public key. After receiving this key, the producer use it to encrypt the content and send it back to the user. The main consequence of such approach is disabling caching mechanism as only one user can decrypt the content. 

\descr{Broadcast encryption~\cite{broadcast_enc0,broadcast_enc}}  allows a ``broadcaster'' to send an encrypted message to a set of receivers $n$, each of which has a different private key. Given any subset of $n$, the broadcaster can construct an encrypted message so that only the receivers in the subset can decrypt it. Using broadcast encryption to guarantee confidentiality in CON presents several advantages. First, the publisher of a content can encrypt it only once, for a known subset of users. Also, the publisher can precompute and store, or generate new decryption keys on the fly, for already published content. Further, since encrypted content can be consumed by many users, the benefit from  caching can be preserved in the network. However, the publisher should generate and store as many keys as the number of clients ($n$).
Also, producer's public key and ciphertexts would be of size $O(\sqrt{n})$~\cite{broadcast_enc}, which may result into a significant communication overhead.

\descr{Proxy re-encryption~\cite{proxy_enc}}  allows a third-party (called proxy) to (re)encrypt a ciphertext which has been encrypted for Alice, so that it can be decrypted by another user, e.g., Bob. The proxy is considered ``semi-trusted" because it does not see the content of the messages being translated. In the CON scenario, 
the content provider could generate a pair of public/private key ($PK_{P}$/$SK_{P}$) for each content object. The content, $m$, is then encrypted as $m_{s} = ENC(M,SK_{P})$. Whenever a client $C$ (with public/private key pair $PK_{C}$/$SK_{C}$) retrieves the content $m_{s}$, it queries the content publisher to generate a re-encryption key by sending $PK_{C}$. $C$ then receives a transformation key $PK_{PC}$ from the publisher that allows him to re-encrypt content $m_{s}$ so that he can decrypt it (i.e., $DEC(RE-ENC(m_{s},PK_{PC}), SK_{C})$ is the original content $m$).
This allows the message to be encrypted only once and lets the producer retain control over the decryption since he can refuse the delivery of the transformation key. Also, key management is simplified as the producer creates transformation keys on the fly and does not have to store any additional keys besides his $PK_{P}$/$SK_{P}$. The encrypted content is disseminated as $m_{s}$ to all users and allows them to benefit from nearby caches. However, it requires both asymmetric encryption and re-encryption (key transformation), which are computationally more expensive than commonly used symmetric-key encryption algorithms. Nonetheless, as the data is encrypted  only once, this overhead can be acceptable in many cases.

\descr{Cover files.} Arianfar et al. \cite{content_priv} described an algorithm to mix legitimate content with so-called ``cover files''.  The content publisher selects a cover content to mix with legitimate content. Cover files are known to both user and the adversary $Adv$. All files are cut in equally-sized blocks and padding is used when necessary. For all $k$ tuples, composed of cover and legitimate blocks, the publisher computes the exclusive-or and publishes the result. For instance, if $k$=2 and given the blocks $c_{1}$, $c_{2}$, $l_{1}$ and $l_{2}$ ($c$ for cover and $l$ for legitimate), the publisher computes $c_{1} \oplus c_{2}, c_{1} \oplus l_{1}, c_{1} \oplus l_{2}, c_{2} \oplus l_{1}, c_{2} \oplus l_{2}$ and $l_{1} \oplus l_{2}$ and publishes them. 
Using a secure side channel, the user retrieves meta information, such as the content hash and length in blocks, cover blocks and the algorithm for generating the names of each block. To retrieve the content, the user requests chunks per name and uses belief propagation or Gaussian elimination to reconstruct the original file.   
However, this technique requires the cooperation of producers who have to generate large amounts of data and xor them. In fact, as pointed out in~\cite{content_priv}, the publisher must produce all the chunks in advance and thus, must perform $O((\alpha+\beta)^{k})$ operations where $\alpha$ represents the number of legitimate blocks, $\beta$ the number of cover blocks, and $k$ the possible number of permutations. Hence, this method incurs several drawbacks that make it unpractical for real-world use.

\subsection{Name privacy}
\label{sec:name}
Name privacy arises from the semantic correlation between human-readable content name and the content itself. Unlike IP, where addresses represent hosts in the network and are not semantically correlated with the content, CCNx~\cite{ccnx} names the content itself and routes data based on content names. Unfortunately, such a property creates an imminent privacy threat as the content names are not only visible but also expected to be semantically related to the content itself (e.g., /US/WebMD/AIDS/Symptoms/html). Although this appears similar to an HTTP connection over IP, it is actually more fundamental in CCNx, as content names cannot be encrypted like the URLs in HTTPS connections. Unlike CCNx, some other architectural proposals such as \cite{dona, netinf} use flat names that are not human-readable. However, all of these proposals rely on a Name Resolution Service (NRS) that performs the translation from human-readable names. Since NRS information is public and accessible to an adversary, even architectures with non human-readable names create the same inherent threat, due to the naming of content pieces.

\subsubsection*{Potential Solutions}

\descr{Bloom filters.} The main name privacy challenge in CONs is keeping the name private while ensuring accessibility and routability. A possible solution is to use Bloom filters~\cite{bloomSurvey} %
to identify content. The resulting architecture would be composed of three main blocks: 
\begin{compactenum}
\vspace{0.1cm}
\item A hierarchical bloom filter used as the routing table. %
\vspace{0.1cm}
\item A counting bloom filter for each interface used as a PIT table \cite{dipit}.
\vspace{0.1cm}
\item A hierarchical bloom filter used as the router storage.
\vspace{0.1cm}
\end{compactenum}

\begin{figure}[t!]
  \begin{center}
  \includegraphics[scale=0.55]{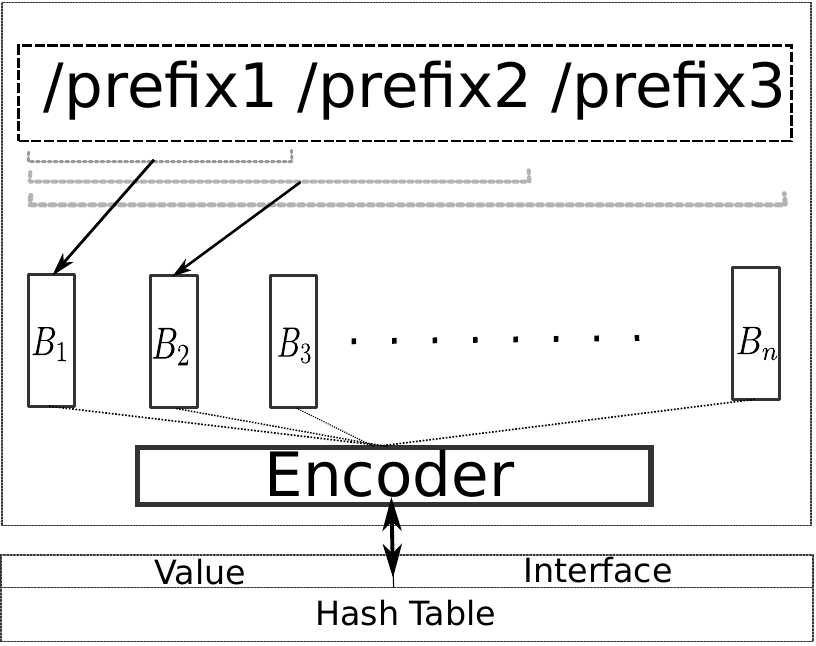}
  \caption{Routing Table using hierarchical bloom filter.}
  \label{fig:bf}
  \vspace{-0.5cm}
	\end{center}	
\end{figure}

Fig.~\ref{fig:bf} shows an example architecture for a routing table. Rather than sending the request in the clear for a hierarchically named content, a client would compute the 
corresponding hierarchical bloom filter as $HB=(B_{1}, B_{2},..., B_{n})$ where $B_{i}$ is the bloom filter of name components up to the $i$-th component. For instance, when asking for \textit{/NYtimes/article/green-econmy}, the client computes a bloom filter $B_{1}$ of \textit{/NYtimes/}, $B_{2}$ of \textit{/NYtimes/article} and
$B_{3}$ of \textit{/NYtimes/article/green-econmy}.  
This scheme would work on CCNx as follows: a router checks whether the last filter ($B_{n}$ since it contains the exact content name) is in its content store, if so, the content is returned to the customer. If not, the router verifies whether  
$B_{n}$ exists in any PIT table. If a match is found, the counting Bloom filter of the corresponding PIT is updated (add one) and the interest is dropped since a request has already been forwarded. Finally, if the interest is not available in the content store nor in any PIT table, the routing process starts: the router checks for longest prefix match on its routing table by starting from $B_{n}$ going all the way to $B_{1}$ until it finds a matching routing entry. 
This approach would enjoy from the obfuscation of content name resulting from transforming it into a random looking string of bits. Also, it can reduce the size needed for storing PIT tables \cite{dipit}, content storage, and routing table, depending on the parameters for the filters and the size of the content name domain. %
However, Bloom filters introduce false positives and periodically require resetting.%

\subsection{Signature privacy}
\label{sec:signature}
One of the main goals of CONs is to decouple content from its location and allow retrieving from nearby caches. 
In order to trust fetched data, some CON architectures, such as, CCNx~\cite{ccnx}, use digital signatures to provide guarantees on provenance and integrity. Although signatures are powerful tools that bind content to its producer, ordinary digital signatures may leak sensitive identity information about the signer.\footnote{\scriptsize CCNx currently offers two choices, RSA and ECDSA, as signature algorithms.}
This is problematic, especially when considering censorship and monitoring, as content from certain publishers can be easily and conveniently identified from the signing key, which is explicitly stated at every data packet in CCNx. %

\subsubsection*{Potential Solutions}

\descr{Confirmer signatures.} A first approach to prevent an adversary from verifying a signature would be to use confirmer signatures. %
Confirmer signatures are undeniable signatures~\cite{deniable} where the signer delegates the verification to a third party (the confirmer), thus, signatures cannot be verified without interacting with this party. Using this method, multiple producers may delegate the verification to a third party and increase the anonymity set for publishers. Although this method would be easy to implement \cite{deniable, deniable2} and would not require any modification to the CCNx schema, it requires a third party as a confirmer and introduces an another round of communication  for signature verification.

\descr{Group signatures~\cite{chaum}} allow the signer to hide in a set of potential signers, thus, providing \textit{signer-ambiguity}. As such, the client can verify that the signature was generated by a member of the group but is unable to tell whom. For instance, a company with multiple employees may use a group signature so that a signature cannot be publicly tracked back to a single user but only to the company. Group signature is efficient since the size of the signature does not depend on the size of the group. However, it assumes the presence of a trusted group manager admitting group members, distributing keys, and revoking anonymity within the group, thus, making it appropriate only in limited settings with collaborating users.

\descr{Ring signatures~\cite{ring}.} As cooperation between users is not always achievable, ring signatures \cite{ring} simplify group signatures by removing both group manager and members interaction. Therefore, there is no need to prearrange group of users, nor
for special procedures for group management and key distribution. Also, the anonymity of the actual signer is always protected. For instance, a company $X$ could collect the public keys of $n$ other trusted companies $Y_{1}, Y_{2}, ..., Y_{n}$, as these keys are publicly available. Then, $X$ can generate a signature $\sigma$ for content $m$ that keeps the signer unidentifiable among  $X$ and other trusted companies. When a client fetches a content $m$, he is able to verify that the content was produced by one of $ \{X\} \bigcup \{Y_{1}, Y_{2}, ..., Y_{n}\}$ without knowing which one it is. This schema allows customers to trust content as long as all possible signers are trustworthy, while making anyone observing their traffic unable to singularly discern the signer. However, the communication overhead introduced by ring signatures is linear in the size of the ring. Also, it would still be possible to enforce censorship based on signatures by blocking all content that list certain entities as one of potential signers.

\descr{Ephemeral identities.} Any content producer can create ephemeral keys to sign content. This would effectively prevent identifying the publisher of content by looking at its signature. However, this would also prevent customers from verifying the source/publisher of a content without an additional mechanism to authenticate it. Luckily, CCNx allows creating unforgeable links by including the hash of a content object in the link and allows these links to be signed. This would allow publishing content signed with ephemeral keys that are not traceable to a long-lived identity, but would still allow users to establish transitive trust when they fetched it, by following link that is published by a trusted party.  For instance, a link to a sensitive content might be published and signed by a trusted blog ($m_{s}$) but the actual content ($m_{a}$) might be published and signed with a one-time indentity and be served through an anonymous hosting service (e.g., rapidshare.com). Any user trusting the blog author can trust $m_{s}$, and thus ($m_{a}$), but an eavesdropper observing $m_{a}$ is unable to link it to its publisher. This approach is very easy to deploy and does not need any modification of the current architecture, however, it cannot hide the access to the link and prevent leakage through its signature.

\section{The Potential of CON Privacy}
\label{sec:Discussions}

\longver{As the amount and the sensitivity of personal information disseminated on the Web increase, 
so might related privacy concerns: according to the Pew Internet \& American Life Project, 
Internet privacy is actually a growing concern among Americans~\cite{pew-privacy}, and
especially in the mobile environment~\cite{pew-mobile}.}

In this section, we compare CON to today's Internet in the context of
a few privacy concepts, such as, anonymity, censoring, traceability, and confidentiality. 
\shortver{(A more detailed discussion is included in the full version of this paper~\cite{full}.)}%

\longver{
\subsection{Anonymity}
\label{sec:discussion_anonymity}
Anonymity on the Internet describes the state of being not identifiable within a set of subjects -- i.e., actions users carry out on the Web cannot be connected with their offline identities. Online anonymity is motivated by a number of factors, as
users are often concerned about harassment, or even threats to their lives, resulting from online activities, such as, protesting and whistle-blowing~\cite{eff}. 
}

\shortver{\descr{Anonymity.} }%
A few techniques are possible today to provide anonymous communications. One na\"{i}ve solution is to rely on a trusted anonymizing proxy 
relaying traffic while removing 
\shortver{identifying information~\cite{Boyan97theanonymizer}.}
\longver{identifying information (such as, IP addresses and cookies) -- see, e.g., the Anonymizer \cite{Boyan97theanonymizer} and the Lucent Personalized Web Assistant \cite{LPWA}.}
However, this introduces a single point of failure and trust, thus, proxy-less techniques have been proposed, 
\shortver{relying on {\em mix networks}~\cite{chaum1981untraceable}. 
Tor~\cite{Tor} is the best-known, and most widely used, low-latency anonymizing tool. Using onion routing and layered encryption, Tor
builds a multi-hop circuit, composed of at least three random nodes.}
\longver{They are usually classified 
based on the application constraints: {\em delay-tolerant} (e.g., email and file-sharing) or {\em low-latency} (e.g., web browsing). 
While both approaches rely on {\em mix networks}, low-latency anonymizing networks cannot afford traffic delaying and reordering as well as the introduction of decoy traffic due to latency exigence. Tor~\cite{Tor} is the best-known, and most widely used, low-latency anonymizing tool. Using onion routing and layered encryption, it builds a multi-hop circuit, composed of at least three random nodes chosen from a central directory, from the user to the Web destination.
}

In CON, proxy-based anonymity could be obtained without the need for an external entity: CON architectures would actually provide, natively, the removal (or substitution) of source and
destination addresses. A neighboring CON router could actually be seen as an anonymizing proxy.
\changed{In reality, however, a local  active adversary could monitor all
connectivity (see Section~\ref{sec:CachePrivacy})}. Regardless, the research community has already started
investigating whether or not onion routing-like techniques can be used in CON: a first answer has been provided by AND$\bar{a}$NA~\cite{andana}, a Tor-like low-latency anonymizing tool for CCN~\cite{ccn}. 
\shortver{Compared to Tor, it only requires two hops and it is resilient against ``hijacking'' attacks (i.e., attacks where the adversary hijacks and modifies server's answers e.g.,~\cite{torAttack}) since data is signed.  However, due to data encryption, CON caching mechanism could not be used.  
Also, AND$\bar{a}$NA inherits some of Tor weaknesses, e.g., the difficulty  of circumventing censorship while retrieving the directory with all participating nodes.}%
\longver{Compared to Tor, it only requires two hops, thus, it is reportedly 2.3 to 7 times faster than Tor when downloading small to medium size files.  
Moreover, since data in CON is signed, attacks where the adversary hijacks and modifies server's answers to de-anonymize user (see, e.g.,~\cite{torAttack}) would not be feasible.
}
Moreover, the portability of AND$\bar{a}$NA to architectures other than CCN depends on the routing protocol. For instance, PSIRP~\cite{psirp} routers do not store any
routing information and rely on a Forwarding Identifier (FI)\footnote{\scriptsize FI is a Bloom filter-based technique used by routers to select the forwarding interface.} provided by the client to route back the content, 
thus, guaranteeing anonymity is very challenging since the routing protocol itself leaks information. 
As the FI carries information about the user, it should be anonymized while ensuring that data is correctly forwarded.  
Whereas, in Dona~\cite{dona} and NetInf~\cite{netinf}, the routing protocol is similar to that used in CCN~\cite{ccn},
hence, it is safe to assume that AND$\bar{a}$NA~\cite{andana} can be used on top of them to provide anonymous communications. 

\longver{However, due to data encryption used in AND$\bar{a}$NA, CON caching mechanism cannot be fully used.  
Also,  AND$\bar{a}$NA inherits some of Tor weaknesses, e.g., the difficulty of circumventing censorship while retrieving the directory with all participating nodes.
In fact, while today Tor Bridges are used as an alternative way to reach the network with alternating results~\cite{GFC,dingledine2011tor}, it is not clear how accomplish this in AND$\bar{a}$NA/CON.

Nonetheless, AND$\bar{a}$NA is actually more resilient to website fingerprinting~\cite{torAttackWeb}, i.e., to
an attacker using patterns such as, time, quantity, and direction of traffic to classify traffic despite encryption and/or tunneling,
thanks to caching, at least against timing. %
Furthermore, CON is not affected by response hijacking attacks~\cite{torAttack}, where the adversary can leverage the lack of data authenticity to modify server's answers (for 
instance,~\cite{torAttack} shows that, by hijacking answers of BitTorrent tracker, an attacker controlling a Tor exit node can successfully de-anonymize users). This is not possible in CON since content is signed (thus, it cannot be altered/modified).
}

\longver{\subsection{Censorship Resistance}

The increasingly important role played by the Web has led many governments to grow their attention in monitoring and censoring Internet traffic. Terms like ``Internet censorship'' are used to denote
control and suppression of the access to, or the publishing of, information on the Internet.

A number of techniques are typically employed to filter and/or block access to certain contents, including
DNS tampering, IP blocking, and keyword filtering, as well as monitoring the usage of specific protocols.
DNS tampering consists in ``deregistering'' the targeted domain, thus, making the domain unreachable. Whereas, IP blocking relies on 
predefined blacklists including IPs that are banned, and keyword filtering leverages Deep Packet Inspection (DPI) 
to dig into traffic and drop packets containing sensitive keywords. 
In today's Internet, both DNS tampering and IP blocking can be easily circumvented. The former can be bypassed by using
public DNS server (e.g., OpenDNS and DNSCrypt~\cite{dnscrypt}) as the main resolver rather than using the default ISP's DNS. 
The latter can also be circumvented using a proxy.
DPI-based filtering is harder to counter and is used increasingly often -- see, e.g., the deployment of China's Great Firewall (GFC)~\cite{china_firewall,GFC}.
  
Unfortunately, }%
\shortver{\descr{Censorship.}} 
Internet censorship appears to be easier in CON architectures. First, and foremost, naming content facilitates keyword filtering. Then, as CON routers have bigger computational and memory resources, content blocking could be carried more effectively, without the need for (expensive) dedicated hardware. Finally, data-monitoring is easier since both interests and data are not encrypted.  Therefore, an attacker only needs to modify the routing protocol so that any ``unwanted'' interest is dropped. 
\longver{Content can be censored independently of its provider and selectively, following a fine-grained censorship approach.

Also, note that DPI techniques can be stateful (i.e., keeping track of the network connections) and, as such, they require a significant amount of memory and processing resource. This requirement impacts filtering capabilities for ``busy'' Internet traffic. For instance, the work in~\cite{ccs_ConceptDoppler} shows that censored Internet traffic in China has diurnal patterns: filtering becomes less effective and lets more than one fourth of the offending packets through during the busy Internet traffic periods. 
By contrast, CON, which adopts the publish/subscribe paradigm, makes the decision to censor content based either on the interest request or on received data, without requiring any external information. 

}
However, if CON traffic is exchanged using AND$\bar{a}$NA~\cite{andana}, there would be a couple of features making it harder for the attacker to censor content. First, both interests and content are encrypted, and, second, as the exit node is usually out of  the attacker's control, the latter cannot delete content. Nonetheless, the effectiveness of AND$\bar{a}$NA (or other techniques in CON) to counter Internet censorship remain an open question that calls for further research.

\longver{
\subsection{Untraceability}
Internet users today are often tracked and profiled. Large amounts of data, corresponding to several different events,
are being collected and mined, so that this knowledge can be used to provide personalized services and behavioral advertising. 
At the same time, privacy advocates have labeled this practice as privacy-invasive, since it allows
companies and providers to build detailed user profiles including personal and sensitive attributes, such as,
sexual orientation, medical conditions, etc.\footnote{\scriptsize See, for instance, a recent incident where targeted advertisement
led to the disclosure of a pregnancy otherwise kept hidden~\cite{target}.}

}

\shortver{\descr{Traceability.} }%
Today, a number of techniques are employed to track users, the most widespread of which are cookies. While it seems straightforward to port cookies to CON, this is not without obstacles. In fact, while in current architectures browsers automatically send cookies to Web servers when fetching data, following so-called {\em same-origin policy}, it is not clear how cookies will be implemented for static content in CON, since data can be fetched from anywhere. Cookies could be transmitted to the source only when fetching dynamic data and, as such, cookie-based tracking mechanisms in CON will be less aggressive as only dynamic content can be tracked. Similar arguments apply for Javascript-based tracking, Supercookies, and Evercookies. 
\longver{Observe that protective measures against profiling are widely available and can be used both with today's Internet and, in the future, in CON. For instance, a few browser add-ons, such as Ghostery~\cite{Ghostery} and DoNotTrack~\cite{DNT} could prevent the browser from sending cookies. }%
However, more aggressive tracking techniques have recently emerged. For instance, {\em Stateless} tracking uses both user IP address and browser fingerprint to uniquely track users on the web \cite{NDSS12Abadi}. 
Mitigating browser fingerprinting can be achieved by using plug-ins, e.g., NoScripts, however, it is still hard to hide user's IP address, unless using techniques for anonymous
communications. However, as CON architectures remove, by design, both parties' identifiers (i.e., source and destination addresses), it would be harder to actually implement IP-based tracking.
The lack of traceability might improve user privacy, however, it naturally raises both security and economic challenges. The former are related to the lack of source address:
for instance, following a security incident (e.g. DoS attacks), this information is often crucial to identify/counter the attacker.
\longver{Also, removing source addresses may thwart security solutions used in firewalls and IDS/IPS tools: e.g., a simple method used today to avoid brute-forcing password guessing attack is to blacklist a host (based on its IP) after a few wrong attempts. }%
Whereas, economic challenges stem from changes that CON might impose on the advertisement model:  ads are usually delivered based on website popularity and user location
and estimating website popularity based on the number of visits is ineffective in CON as caching will ``hide'' a significant amount of traffic. 

Finally, observe that, as discussed in Section~\ref{sec:PrivacyChallenges}, a few CON network components \shortver{(including the role played by neighboring routers)} can actually be used to better track users.
\longver{Specifically, neighboring routers have a significant role in protecting user privacy: if compromised or misused, they can severely endanger it. For instance, a router might collect content, content names, or content signatures to track users' navigation and build accurate profiles.}

\longver{\subsection{Data authenticity and confidentiality} 
One of the main CON's ``selling points'' is that security is built into the data itself, by enforcing content signature. 
Therefore, integrity, provenance, and trustworthiness of content become built-in features. As keys can be treated as named CON data, key distribution does not constitute a major issue.}%
\shortver{\descr{Data authenticity and confidentiality.} } 
While today's Internet requires a ``one-size-fits-all'' trust model, trust in CON is end-to-end,
between data producer and data consumer, and does not depend on any physical or temporal frame. This modularity has two main advantages. (1) Different consumers may easily implement different levels of security, and (2) on CON, one can employ both widely accepted and new trust management models as data is independent from the deployed model.  
However, these features also prompt some challenges. First, we need to identify a set of usable trust mechanisms that can be deployed and used by most users. Second, as all content is signed, it is crucial to assess (and potentially improve) efficiency of signature generation, transmission, verification, and possibly storage. 
\longver{Note that encryption in CON is not applied to publicly available content, thus, creating problems related to data confidentiality. One main disadvantage of current approaches to data encryption (e.g., TLS) is that it inhibits caching, thus, defeating one of the major advantages to improve network performance. TLS seems to be in contradiction with the CON design as: (1) trust is linked to a session and not to the content itself, and (2) only one user can decrypt content, thus, inhibiting caching mechanism. However, as discussed in Section~\ref{sec:content}, solutions like proxy re-encryption might mitigate this issue. 

As a result, }%
\shortver{Finally,} we believe that providing data confidentiality while keeping caching mechanism is one of the major open challenges in CON. 
\longver{Although we have proposed several countermeasures, most of these rely on public key cryptography, so real-world  performance overhead imposed on both clients and servers needs to be thoroughly evaluated. }

\section{Related work}\label{sec:related}
\label{sec:related}
Propelled by the increasing interest for next-generation Internet architectures and, in particular,
Content-Oriented Networking (CON),
the research community has produced a large body of work dealing with CON
building blocks~\cite{ccn,dona,psirp,netinf,triad}, performance~\cite{dipit,caching_perf,wave,probabilistic_caching}, and scalability~\cite{caching_perf02,distributed_caching}.
However, the quest for analyzing and enhancing security in CON is only at the beginning --
in particular, very little work has focused on privacy and anonymity.
In this section, we review relevant prior work.

\descr{Security in CON.} 
Wong and Nikander~\cite{name_priv02} address security of naming mechanisms
by constructing content name as the concatenation of content provider's ID, cryptographic ID of the content and some meta-data. Dannewitz et al.~\cite{name_priv01} adopt a similar approach where content name is defined as the concatenation of the hash of the public key and a set of attributes. Both schemes rely on cryptographic hash functions to name the content, which results in a human-unreadable flat naming. Smetters et al.~\cite{pracName} show that these schemes have several drawbacks, including the need for an indirection mechanism to map and the lack of binding between name and producer's identity. To resolve these shortcomings, they propose to keep hierarchical human readable names while signing both content name and the content itself, using producer's public key. %
Gasti et al.~\cite{gasti2012ddos} study DoS and DDOS in CCN~\cite{ccn} by presenting attacks and proposing some initial countermeasures. 
In another context, Burke et al.~\cite{burke} propose a secure lighting systems over Named-Data Networking (NDN), 
providing control access to fixtures via authorization policies, coupled with strong authentication. 
\longver{This approach is a first attempt to port CON out of the content distribution scenario.}

\descr{Privacy Issues in CON.}
To the best of our knowledge, the only related privacy study is the recent article by Lauinger et al. in~\cite{tobi_2,tobi_1},
that covers security and privacy issues of CCN~\cite{ccn}. Specifically, they highlight a few Denial-of-Service (DoS) vulnerabilities as well as different cache-related attacks.  
\longver{
In CCN, a possible DoS attack (also discussed in~\cite{ccn}) relies on resource exhaustion, targeting either routers or content source. Routers are forced to perform expensive computation, such as, signature verification, which negatively affects the quality of service and can ultimately block traffic. Content source can also be flooded with a huge number of interests, ending up denying service to legitimate users. Additional DoS attacks mainly target the cache mechanism, to either decrease network performance or to gain free and uncontrolled storage. Transforming the cache into a permanent storage  is achieved by continuously issuing interests for a desired file. Decreased network performance can also be achieved through cache pollution.

}%
From the privacy perspective, work in~\cite{tobi_2,tobi_1} identifies the issue of information leakage through caches in CCN.
It proposes a few simple countermeasures, following \textit{detection} and \textit{prevention} approaches. The former can be achieved using techniques similar to those addressing cache pollution attacks  in IP~\cite{pollution}, although such an approach can be difficult to port to CON due to the lack of source address. The latter can actually be global, i.e., treating all traffic as sensitive, delaying all traffic, or deploying a shared cache to circumvent the attack. Alternatively, a selective prevention approach may try to distinguish between sensitive and non-sensitive content, based on content popularity and context (time, location), and then delay or tunnel sensitive content. 
\longver{
It is not clear, however, how to implement the selection mechanism to distinguish between private and non-private content, but authors of~\cite{tobi_1} suggest to implement this service either in the network layer (i.e., the router classifies the content) or by the host (i.e., content source tags sensitive content). %
Such classification is in turn a very challenging task, since privacy is a relative notion that changes from one user to another. Also, \textit{censorship and surveillance} are briefly discussed, although no countermeasures besides tunneling have been proposed.

}%
Our work extends that in~\cite{tobi_2,tobi_1} by encompassing {\em all} privacy aspects: caching, naming, signature, and content. Also, it is more general as it does not only consider CCN~\cite{ccn}, but CON in general, independently of the
specific instantiation. Furthermore, when suggesting countermeasures, we only propose techniques that can be applied with a minimal change to the architecture. 

\descr{Anonymity in CON.}
AND$\bar{a}$NA~\cite{andana} proposes a Tor-like anonymizing tool for CCN~\cite{ccn} to provide provable anonymity. It also aims to privacy protection via simple tunneling. However, 
\longver{
as discussed in Section~\ref{sec:discussion_anonymity},
} AND$\bar{a}$NA is an ``all-in-one'' solution that introduces latency and impedes caching. Whereas, fine-grained privacy solutions are needed, since a widespread use of tunneling would inherently take away most of
CON benefits in terms of performance and scalability.
\longver{
To provide censorship resistance, Arianfar et al. \cite{content_priv} describes an algorithm to mix legitimate sensitive content with so-called ``cover files'' to hide it. By monitoring the content, an adversary would only see the ``mixed'' content, which prevents him from censoring the content.  
}%

\section{Conclusion}\label{sec:conclusion}
\label{sec:conclusion}
Content-Oriented Networking (CON) proposes a major transition from today's Internet to a new content-based architecture. This radical change calls for a thorough analysis of both security and privacy guarantees. 
CON comes with a potential benefit to security, including a security-by-design
approach based on digital signatures that provides data integrity and origin authentication, as well as  trust support. 
However, prior to our work, it remained an open question whether or not, and to what 
extent, this emerging networking paradigm bears new privacy threats. 

This paper presented a first-of-its-kind, systematic analysis of privacy issues in CON as a generic paradigm, 
discussing different attacks and detailing their impact on user privacy.
We also proposed several countermeasures while attempting to balance the trade-off between privacy, performance, 
and changes to the architecture.
In the process, we identified a number of interesting research challenges that call for further work in the area.

\changed{Naturally, our work does not end here: first, and foremost, we are working toward further evaluating 
the feasibility of our proposed countermeasures and their effective deployment; next, we plan to provide
an in-depth study of multiple encryption and signature techniques and their impact on network performance; finally, we intend to analyze the impact of privacy-enhancing and CON-native technologies on Web economy and advertisement models.}

\balance

\bibliographystyle{abbrv}
\bibliography{bibfile}

\begin{thebibliography}{10}

\bibitem{acs}
G.~Acs, M.~Conti, P.~Gasti, C.~Ghali, and G.~Tsudik.
\newblock {Cache Privacy in Named-Data Networking}.
\newblock In {\em ICDCS}, 2013.

\bibitem{internationalNaming}
W.~Adjie-Winoto, E.~Schwartz, H.~Balakrishnan, and J.~Lilley.
\newblock The design and implementation of an intentional naming system.
\newblock {\em ACM SIGOPS Operating Systems Review}, 34(2), 2000.

\bibitem{survey_ccn_02}
B.~Ahlgren, C.~Dannewitz, C.~Imbrenda, D.~Kutscher, and B.~Ohlman.
\newblock {A survey of Information-Centric Networking}.
\newblock {\em IEEE Communications Magazine}, 50(7), 2012.

\bibitem{PersoNamespaces}
M.~Allman.
\newblock Personal namespaces.
\newblock In {\em HotNets}, 2007.

\bibitem{netinf}
M.~Ambrosio, M.~Marchisio, V.~Vercellone, and et~al.
\newblock {Second NetInf Architecture Description}.
\newblock 4WARD Deliverable D6.2,
  \url{http://www.4ward-project.eu/index.php?id=192}, 2010.

\bibitem{aggregation_effect}
J.~Ardelius, B.~Gr\"{o}nvall, L.~Westberg, and A.~Arvidsson.
\newblock On the effects of caching in access aggregation networks.
\newblock In {\em ICN}, 2012.

\bibitem{content_priv}
S.~Arianfar, T.~Koponen, B.~Raghavan, and S.~Shenker.
\newblock {On preserving privacy in Content-Oriented Networks}.
\newblock In {\em ICN}, 2011.

\bibitem{layered-naming}
H.~Balakrishnan, K.~Lakshminarayanan, S.~Ratnasamy, S.~Shenker, I.~Stoica, and
  M.~Walfish.
\newblock {A Layered Naming Architecture for the Internet}.
\newblock In {\em SIGCOMM}, 2004.

\bibitem{LPWA}
{Bell Labs}.
\newblock {The Lucent Personalized Web Assistant}.
\newblock \url{http://www.bell-labs.com/project/lpwa/}, 1998.

\bibitem{dpi}
R.~Bendrath and M.~Mueller.
\newblock {The End of the Net as We Know It? Deep Packet Inspection and
  Internet Governance}.
\newblock {\em New Media and Society}, 13(7), 2011.

\bibitem{proxy_enc}
M.~Blaze, G.~Bleumer, and M.~Strauss.
\newblock Divertible protocols and atomic proxy cryptography.
\newblock In {\em EUROCRYPT}, 1998.

\bibitem{broadcast_enc}
D.~Boneh, C.~Gentry, and B.~Waters.
\newblock {Collusion-resistant Broadcast Encryption with Short Ciphertexts and
  Private Keys}.
\newblock In {\em CRYPTO}, 2005.

\bibitem{distributed_caching}
S.~Borst, V.~Gupta, and A.~Walid.
\newblock {Distributed Caching Algorithms for Content Distribution Networks}.
\newblock In {\em INFOCOM}, 2010.

\bibitem{Boyan97theanonymizer}
J.~Boyan.
\newblock {The Anonymizer -- Protecting User Privacy on the Web}, 1997.

\bibitem{bloomSurvey}
A.~Broder, M.~Mitzenmacher, and A.~Broder.
\newblock {Network Applications of Bloom Filters: A Survey}.
\newblock {\em Internet Mathematics}, 1, 2002.

\bibitem{burke}
J.~Burke, P.~Gasti, N.~Nathan, and G.~Tsudik.
\newblock {Securing Instrumented Environments over Content-Centric Networking:
  the Case of Lighting Control}.
\newblock In {\em NOMEN}, 2013.

\bibitem{flatLabel}
M.~Caesar, T.~Condie, J.~Kannan, K.~Lakshminarayanan, and I.~Stoica.
\newblock {ROFL: Routing on Flat Labels}.
\newblock In {\em SIGCOMM}, 2006.

\bibitem{cbcb}
A.~Carzaniga, M.~J. Rutherford, and A.~L. Wolf.
\newblock {A Routing Scheme for Content-Based Networking}.
\newblock In {\em INFOCOM}, 2004.

\bibitem{hmip}
C.~Castelluccia.
\newblock {HMIPv6: A Hierarchical Mobile IPv6 Proposal}.
\newblock {\em SIGMOBILE Mobile Computing and Communications Review}, 4(1),
  2000.

\bibitem{deniable}
D.~Chaum.
\newblock Desinated-confirmer signature systems, 1994.
\newblock US Patent 5,373,558.

\bibitem{chaum}
D.~Chaum and E.~Van~Heyst.
\newblock Group signatures.
\newblock In {\em EUROCRYPT}, 1991.

\bibitem{cho2008content}
K.~Cho, J.~Choi, D.~Ko, T.~Kwon, and Y.~Choi.
\newblock {Content-oriented networking as a future internet infrastructure:
  Concepts, strengths, and application scenarios}.
\newblock In {\em Future Internet Technologies}, 2008.

\bibitem{wave}
K.~Cho, M.~Lee, K.~Park, T.~Kwon, Y.~Choi, and S.~Pack.
\newblock {WAVE: Popularity-based and collaborative in-network caching for
  Content-Oriented Networks}.
\newblock In {\em INFOCOM Workshops}, 2012.

\bibitem{survey_ccn_01}
J.~Choi, J.~Han, E.~Cho, T.~Kwon, and Y.~Choi.
\newblock {A Survey on Content-Oriented Networking for Efficient Content
  Delivery}.
\newblock {\em IEEE Communications Magazine}, 49(3), 2011.

\bibitem{cisco}
{Cisco, Inc.}
\newblock {Cisco Visual Networking Index: Forecast and Methodology,
  2011--2016}.
\newblock \url{http://preview.tinyurl.com/3p7v28}, 2012.

\bibitem{mobile_ip}
X.~P. Costa and H.~Hartenstein.
\newblock A simulation study on the performance of mobile ipv6 in a wlan-based
  cellular network.
\newblock {\em Computer Networks}, 40(1), 2002.

\bibitem{ccs_ConceptDoppler}
J.~R. Crandall and E.~Barr.
\newblock Conceptdoppler: A weather tracker for internet censorship.
\newblock In {\em CCS}, 2007.

\bibitem{china_firewall}
J.~R. Crandall, D.~Zinn, M.~Byrd, E.~T. Barr, and R.~East.
\newblock Conceptdoppler: a weather tracker for internet censorship.
\newblock In {\em CCS}, 2007.

\bibitem{name_priv01}
C.~Dannewitz, J.~Golic, B.~Ohlman, and B.~Ahlgren.
\newblock {Secure Naming for a Network of Information}.
\newblock In {\em INFOCOM Workshops}, 2010.

\bibitem{pollution}
L.~Deng, Y.~Gao, Y.~Chen, and A.~Kuzmanovic.
\newblock Pollution attacks and defenses for internet caching systems.
\newblock {\em Comput. Netw.}, 2008.

\bibitem{andana}
S.~DiBenedetto, P.~Gasti, G.~Tsudik, and E.~Uzun.
\newblock Andana: Anonymous named data networking application.
\newblock In {\em NDSS}, 2012.

\bibitem{dingledine2011tor}
R.~Dingledine.
\newblock {Tor and circumvention: Lessons learned (Invited Talk)}.
\newblock In {\em CRYPTO}, 2011.

\bibitem{Tor}
R.~Dingledine, N.~Mathewson, and P.~Syverson.
\newblock {Tor: The Second-Generation Onion Router}.
\newblock In {\em Usenix Security}, 2004.

\bibitem{ssl_flow}
T.~Duong and J.~Rizzo.
\newblock Here come the $\oplus$ ninjas.
\newblock In {\em Ekoparty Security Conference}, 2011.

\bibitem{eff}
{Electronic Frontier Foundation}.
\newblock {Anonymity}.
\newblock \url{https://www.eff.org/issues/anonymity}, 2012.

\bibitem{Ghostery}
{Evidon, Inc.}
\newblock Ghostery.
\newblock \url{http://www.ghostery.com/}, 2012.

\bibitem{timing_attack}
E.~W. Felten and M.~A. Schneider.
\newblock {Timing Attacks on Web Privacy}.
\newblock In {\em CCS}, 2000.

\bibitem{broadcast_enc0}
A.~Fiat and M.~Naor.
\newblock {Broadcast Encryption}.
\newblock In {\em CRYPTO}, 1994.

\bibitem{User-Relative-Names}
B.~Ford, J.~Strauss, C.~Lesniewski-laas, S.~Rhea, F.~Kaashoek, and R.~Morris.
\newblock {User-Relative Names for Globally Connected Personal Devices}.
\newblock In {\em IPTPS}, 2006.

\bibitem{deniable2}
S.~Galbraith and W.~Mao.
\newblock Invisibility and anonymity of undeniable and confirmer signatures.
\newblock In {\em CT-RSA}, 2003.

\bibitem{gasti2012ddos}
P.~Gasti, G.~Tsudik, E.~Uzun, and L.~Zhang.
\newblock {DoS and DDoS in Named-Data Networking}.
\newblock In {\em ICCCN (To Appear)}, 2013.

\bibitem{georgiev2012most}
M.~Georgiev, S.~Iyengar, S.~Jana, R.~Anubhai, D.~Boneh, and V.~Shmatikov.
\newblock {The most dangerous code in the world: validating SSL certificates in
  non-browser software}.
\newblock In {\em CCS}, 2012.

\bibitem{triad}
M.~Gritter and D.~R. Cheriton.
\newblock {An Architecture for Content Routing Support in the Internet}.
\newblock In {\em USITS}, 2001.

\bibitem{target}
K.~Hill.
\newblock {How Target Figured Out A Teen Girl Was Pregnant Before Her Father
  Did}.
\newblock \url{http://preview.tinyurl.com/7jbntx3}, 2012.

\bibitem{ccn}
V.~Jacobson, D.~K. Smetters, J.~D. Thornton, M.~F. Plass, N.~H. Briggs, and
  R.~L. Braynard.
\newblock {Networking Named Content}.
\newblock In {\em CoNEXT}, 2009.

\bibitem{ssl_hack}
G.~Keizer.
\newblock {Hackers may have stolen over 200 SSL certificates}.
\newblock
  \url{http://www.computerworld.com/s/article/9219663/Hackers_may_have_stolen_over_200_SSL_certificates},
  2011.

\bibitem{dona}
T.~Koponen, M.~Chawla, B.-G. Chun, A.~Ermolinskiy, K.~H. Kim, S.~Shenker, and
  I.~Stoica.
\newblock {A Data-Oriented (and beyond) Network Architecture}.
\newblock {\em SIGCOMM Computer Communication Review}, 37(4), 2007.

\bibitem{OceanStore}
J.~Kubiatowicz, D.~Bindel, Y.~Chen, S.~Czerwinski, P.~Eaton, D.~Geels,
  R.~Gummadi, S.~Rhea, H.~Weatherspoon, W.~Weimer, C.~Wells, and B.~Zhao.
\newblock Oceanstore: an architecture for global-scale persistent storage.
\newblock {\em SIGPLAN Notes}, 35(11), 2000.

\bibitem{h-cache}
N.~Laoutaris, S.~Syntila, and I.~Stavrakakis.
\newblock {Meta Algorithms for Hierarchical Web Caches}.
\newblock In {\em IPCCC}, 2004.

\bibitem{tobi_2}
T.~Lauinger, N.~Laoutaris, P.~Rodriguez, T.~Strufe, E.~Biersack, and E.~Kirda.
\newblock {Privacy Implications of Ubiquitous Caching in Named Data Networking
  Architectures}.
\newblock Technical report, TR-iSecLab-0812-001, iSecLab, 2012.

\bibitem{tobi_1}
T.~Lauinger, N.~Laoutaris, P.~Rodriguez, T.~Strufe, E.~Biersack, and E.~Kirda.
\newblock Privacy risks in named data networking: what is the cost of
  performance?
\newblock {\em SIGCOMM Computer Communications Review}, October 2012.

\bibitem{bt02}
S.~Le~Blond, A.~Legout, F.~Lefessant, W.~Dabbous, and M.~Kaafar.
\newblock {Spying the World from Your Laptop: Identifying and Profiling Content
  Providers and Big Downloaders in BitTorrent}.
\newblock In {\em LEET}, 2010.

\bibitem{torAttack}
S.~Le~Blond, P.~Manils, A.~Chaabane, M.~A. Kaafar, C.~Castelluccia, A.~Legout,
  and W.~Dabbous.
\newblock {One bad apple spoils the bunch: exploiting P2P applications to trace
  and profile Tor users}.
\newblock In {\em LEET}, 2011.

\bibitem{lisp-dht}
L.~Mathy and L.~Iannone.
\newblock Lisp-dht: towards a dht to map identifiers onto locators.
\newblock In {\em CoNEXT}, 2008.

\bibitem{DNT}
J.~Mayer and A.~Narayanan.
\newblock {Do Not Track, Universal Web Tracking Opt Out}.
\newblock \url{http://donottrack.us/}, 2012.

\bibitem{ndn}
{Named Data Networking (NDN)}.
\newblock \url{http://www.named-data.net/}, 2012.

\bibitem{NSF}
{National Science Foundation}.
\newblock {NSF Future Internet Architecture Project}.
\newblock \url{http://www.nets-fia.net/}, 2010.

\bibitem{dnscrypt}
OpenDNS.
\newblock {Introducing DNSCrypt}.
\newblock \url{http://www.opendns.com/technology/dnscrypt/}.

\bibitem{ccn_interest}
{Palo Alto Research Center, Inc.}
\newblock {Project CCNx: Interest Message}.
\newblock
  \url{http://www.ccnx.org/releases/latest/doc/technical/InterestMessage.html},
  2012.

\bibitem{ccnx}
{Palo Alto Research Center, Inc.}
\newblock {Project CCNx: Open-Source Implementation and Documentation}.
\newblock \url{http://www.ccnx.org/}, 2012.

\bibitem{torAttackWeb}
A.~Panchenko, L.~Niessen, A.~Zinnen, and T.~Engel.
\newblock Website fingerprinting in onion routing based anonymization networks.
\newblock In {\em WPES}, 2011.

\bibitem{pew-privacy}
{Pew Internet \& American Life Project}.
\newblock {Search Engine Use 2012}.
\newblock
  \url{http://pewinternet.org/~/media//Files/Reports/2012/PIP_Search_Engine_Use_2012.pdf},
  2012.

\bibitem{pew-mobile}
{Pew Internet \& American Life Project}.
\newblock {Search Engine Use 2012}.
\newblock
  \url{http://pewinternet.org/~/media//Files/Reports/2012/PIP_MobilePrivacyManagement.pdf},
  2012.

\bibitem{probabilistic_caching}
I.~Psaras, W.~K. Chai, and G.~Pavlou.
\newblock Probabilistic in-network caching for information-centric networks.
\newblock In {\em ICN}, 2012.

\bibitem{sdsi}
R.~L. Rivest and B.~Lampson.
\newblock {SDSI -- A Simple Distributed Security Infrastructure}.
\newblock In {\em CRYPTO}, 1996.

\bibitem{ring}
R.~L. Rivest, A.~Shamir, and Y.~Tauman.
\newblock {How to Leak a Secret}.
\newblock In {\em ASIACRYPT}, 2001.

\bibitem{caching_perf}
G.~Rossini and D.~Rossi.
\newblock A dive into the caching performance of content centric networking.
\newblock In {\em CAMAD}, 2012.

\bibitem{caching_perf02}
G.~Rossini and D.~Rossi.
\newblock {On sizing CCN content stores by exploiting topological information}.
\newblock In {\em NOMEN}, 2012.

\bibitem{bt01}
R.~C. Rum\'{\i}n, N.~Laoutaris, X.~Yang, G.~Siganos, and P.~Rodriguez.
\newblock {Deep Diving into BitTorrent Locality}.
\newblock In {\em INFOCOM}, 2011.

\bibitem{pracName}
D.~Smetters and V.~Jacobson.
\newblock {Securing Network Content}.
\newblock Technical Report,
  \url{www.parc.com/content/attachments/securing-network-content-tr.pdf}, 2009.

\bibitem{psirp}
K.~Visala, D.~Lagutin, and S.~Tarkoma.
\newblock {LANES: An Inter-Domain Data-Oriented Routing Architecture}.
\newblock In {\em ReArch}, 2009.

\bibitem{squid_cache}
D.~Wessels.
\newblock {Configuring Hierarchical Squid Caches}.
\newblock
  \url{http://old.squid-cache.org/Doc/Hierarchy-Tutorial/tutorial-1.html}.

\bibitem{internet_cache}
D.~Wessels and K.~Claffy.
\newblock {Internet Caching Protocol -- RFC2186}.
\newblock \url{http://tools.ietf.org/html/rfc2186}, 1997.

\bibitem{GFC}
P.~Winter and S.~Lindskog.
\newblock {How the Great Firewall of China is Blocking Tor}.
\newblock In {\em FOCI}, 2012.

\bibitem{name_priv02}
W.~Wong and P.~Nikander.
\newblock {Secure Naming in Information-Centric Networks}.
\newblock In {\em ReARCH}, 2010.

\bibitem{NDSS12Abadi}
T.-F. Yen, Y.~Xie, F.~Yu, R.~P. Yu, and M.~Abadi.
\newblock Host fingerprinting and tracking on the web: Privacy and security
  implications.
\newblock In {\em NDSS}, 2012.

\bibitem{dipit}
W.~You, B.~Mathieu, P.~Truong, J.-F. Peltier, and G.~Simon.
\newblock {Realistic Storage of Pending Requests in Content-Centric Network
  Routers}.
\newblock In {\em ICC}, 2012.

\bibitem{tapestry}
B.~Zhao, J.~Kubiatowicz, and A.~Joseph.
\newblock {Tapestry: A Fault-Tolerant Wide-Area Application Infrastructure}.
\newblock {\em SIGCOMM Computer Communication Review}, 32(1), 2002.

\end{thebibliography}

\end{document}